\documentclass[namedreferences]{solarphysics}

\usepackage[optionalrh,showbiblabels]{spr-sola-addons} 
\usepackage{epsfig}          
\usepackage{graphicx, rotating}        
\usepackage{amssymb}        
\usepackage{color}           
\usepackage{breakurl}        
\usepackage{epsfig}
\usepackage{bm} 
\usepackage[T1]{fontenc}      
\usepackage{amsfonts}

\newcommand{\adv}{{\it Adv. Space Res.}} 
 
\newcommand{\aap}{    {\it Astron. Astrophys.}}

\newcommand{\apj}{    {\it Astrophys. J.}}
\newcommand{\apjl}{   {\it Astrophys. J. Lett.}}

\newcommand{\solphys}{{\it Solar Phys.}}

\chardef\us=`\_

\begin{document}

\begin{article}
\begin{opening}

\title{Solar flare forecasting from magnetic feature properties generated by Solar Monitor Active Region Tracker\\ {\it Solar Physics}}

\author[addressref={aff1},corref,email={katarina.domijan@mu.ie}]{\inits{K.}\fnm{Katarina}~\lnm{Domijan}\orcid{0000-0002-4268-2236}}
\author[addressref={aff2}]{\inits{D.S.}\fnm{D.Shaun}~\lnm{Bloomfield}\orcid{0000-0002-4183-9895}}
\author[addressref={aff3}]{\inits{F.} \fnm{Fran\c cois}~\lnm{Piti\'e}\orcid{0000-0003-4599-0549}} 

\address[id=aff1]{Maynooth University}
\address[id=aff2]{Northumbria University}
\address[id=aff3]{Trinity College Dublin}

\runningauthor{Domijan et al.}
\runningtitle{Solar flare forecasting from MF properties generated by SMART}

\begin{abstract}
We study the predictive capabilities of magnetic feature properties (MF) generated by Solar Monitor Active Region Tracker (SMART) \citeauthor{higgins11}(\adv{} \textbf{47}, 2105,
\citeyear{higgins11}) for solar flare forecasting from two datasets: the full dataset of SMART detections from 1996 to 2010 that has been previously studied by \citeauthor{ahmed11} (\solphys{} \textbf{283(1)}, 179,
\citeyear{ahmed11}) and a subset of that dataset which only includes detections that are NOAA active regions (ARs).

Main contributions: we use marginal relevance as a filter feature selection method  to identify most useful SMART MF properties for separating flaring from non-flaring detections and logistic regression to derive classification rules to predict future observations.  For comparison, we employ a Random Forest, Support Vector Machine and a set of Deep Neural Network models, as well as Lasso for feature selection.  Using the linear model with three features we obtain significantly better results (TSS=0.84) to those reported by \citeauthor{ahmed11} (\solphys{} \textbf{283(1)}, 179, \citeyear{ahmed11}) for the full dataset of SMART detections. The same model produced competitive results (TSS=0.67) for the dataset of SMART detections that are NOAA ARs which can be compared to a broader section of flare forecasting literature. We show that more complex models are not required for this data.

\end{abstract}
\end{opening}
-------------------------------------------------

\section{Introduction}\label{sec:intro}

Solar flares strongly influence space weather and their prediction using photospheric magnetic field observations has been studied extensively in recent years (e.g. \cite{abramenko05,mcateer05,schrijver07,leka07,georgoulis07,qahwaji08,colak08,barnesleka08,colak09,mason10,yu10,yang13, alghraibah15, boucheron15,bobra2015, liu17,daei17,  gheibi17, liudeng17, raboonik17, nishizuka2017, nishizuka18, huang18})

In this paper we analysed a dataset of magnetic feature  (MF) properties generated by Solar Monitor Active Region Tracker (SMART) \citep{higgins11}, an automated system for detecting and tracking active regions (AR) from SOHO Michelson Doppler Interferometer (MDI) magnetograms. SMART determines MF properties such as region size, total flux, flux imbalance, flux emergence rate, Schrijver's R-value and Falconer's measurement of non-potentiality. Each MF detection was classed as flaring or non-flaring if it produced a C-class or above flare within the 24 hours following the observation.

This dataset was previously analysed by \cite{ahmed11} and in this paper one of the aims was to improve on their results. We considered a number of classification approaches: binary logistic regression (LR) \citep{cox58}, which is a linear classifier, as well as the classifiers that allow for nonlinear classification rules, namely: Random Forests (RF) \citep{breiman01}, support vector machines (SVMs) \citep{vapnik98} and a set of Deep Feedforward Neural Network (DNN) architectures. We considered feature selection for the linear model: the LR classifier was applied to a small subset of MF properties selected by a marginal relevance (MR) criteria  \citep{dudoit02} and the full feature set. We also used Lasso \citep{tibshirani96}, a model related to LR that simultaneously performs classification and feature selection.

To assess how the results of a predictive model will generalize to an independent dataset we used cross-validation where the training of algorithms and feature selection are carried out on the training set and the presented results are shown for the test set. True Skill Scores (TSS), Heidke Skill Score (HSS), Receiver Operating Characteristic (ROC) curves and Area Under ROC curve (AUC) were used as measures of classifier performance.

For the dataset analysed by \cite{ahmed11} we found that the linear classifier using only the top three features selected by MR yielded good classification rates with the highest TSS of 0.84, sensitivity (recall) of $95\%$ and specificity of $89\%$. This is a significant improvement on the previous analysis of this data. None of the other approaches that we considered exceeded this performance.

SMART detects MFs automatically and independently from NOAA active regions. A large number of detections are small magnetic flux regions that have no associated sunspot structure and do not possess many of the properties that SMART calculates, yielding values close to zero for some of the features. These detections never flare and it is relatively easy for a forecasting system to get them correct. In order to compare our results to a broader section of the flare forecasting literature, we analysed a second set of results that correspond to SMART detections which are NOAA active regions by initially filtering the SMART dataset. For this reduced dataset, the same linear classifier with the top three features selected by MR yielded TSS of 0.67 with corresponding sensitivity and specificity of $87\%$ and $80\%$. None of the other models, including more comprehensive searches of the feature space and nonlinear classifiers, were able to improve on this performance.

Based on the classification results as well as the visualization of the data we show that there is no advantage in including a larger number of features or fitting more complex, non-linear models for these datasets.

For comparison with \cite{ahmed11} we used the same split of the data into training and testing sets. For the full dataset of all SMART detection, the training set is large, comprising of 330,000 instances and as such puts constraints on the choice of classifier methodology. For example, kernel based classifiers such as SVM require the computation of an $n \times n$ dimensional kernel matrix and do not scale well to data where $n$, the number of instances is large. To evade this issue we took the approach of subsampling from the full training set to construct 50 smaller training sets of 400 instances each. The SVMs were trained on these small training sets whereas DNNs involve highly parameterized models and were trained on the full training set. LR and RF were trained on the small subsampled training sets and the full training set. We found that using the entire training set to train the algorithms gives no improvement in test classification rates.

The analysis of the data was done in R, a free software environment for statistical computing and graphics \citep{R17}. The graphical displays were produced using ggplot2 \citep{wickham} and plot3D \citep{soetaert17} packages. The code to reproduce the analysis and graphics can be accessed at \url{https://github.com/domijan/Sola}.

The paper is organized as follows: in Section~\ref{sec:data} we describe the dataset, in Section~\ref{sec:methods} we briefly outline the method used for feature selection; the classification algorithms;  the cross-validation settings for assessing classifier performance and the forecast performance measures. Section~\ref{sec:results} presents the results and in Section~\ref{sec:discussion} we make some concluding comments.

\section{Data}\label{sec:data}

Data are line-of-sight magnetograms from \emph{SOHO}/MDI. Magnetic feature properties were extracted by Solar Monitor Active Region Tracking algorithm \citep{higgins11}. Flares are from \emph{Geostationary Operational Environmental Satellite (GOES)} soft X-ray (1--8\,\AA) flare lists provided by NOAA/SWPC.

SMART detects MFs automatically and independently from NOAA active regions. Following \citep{ahmed11} we defined an ``MF detection'' as an individual SMART MF detected in one MDI magnetogram. Each MF detection was classified as flaring or non-flaring if it produced a C-class or above flare within the 24 hours following the observation. In order to minimize the error caused by projection effects, only MF detections located within 45 deg from solar disc center were considered. The dataset comprises of MF detections generated by SMART from April 1996 - Dec 2010. A list of SMART MF features used in this analysis with descriptions is given in Table~\ref{table:features}.

In this paper we study two datasets: the ``full SMART dataset'' of all MF detections generated by SMART from 1996 April 1 to 2010 December 31 and a ``NOAA AR dataset'' containing of only those SMART MF detections that can be associated with NOAA ARs. The second dataset is derived by retaining only those SMART MF detections whose boundaries encompass the coordinates of one or more NOAA ARs after these were time-rotated to the MDI observation times used by SMART.

\section{Methods}\label{sec:methods}

\subsection{Classification algorithms}\label{sec:class}

\textbf{Logistic regression} is a well established framework for modelling and prediction of data where the response variable of interest is binary. It is a subset of the Generalized Linear Models (GLMs)\citep{NeldWedd72} that are widely used across a range of scientific disciplines and are available in almost all statistical software packages.

For each MF detection in a training dataset, we have a feature vector $\bm{X}_{i}$ and an observed class label $Y_i \in \{0,1\}$, denoting if it produced a flare. The distribution of of $Y_i$ is modeled by a Bernoulli ($p_i$) distribution, where $p_i = P(Y_i = 1 |\bm{X}_{i}, \bm{\beta})$  denotes the probability of flare  and $\bm{\beta}$ is the parameter vector. In this model we use the logit link, where  $p_i$ is the logistic function of a linear combination of the explanatory features:

 $$p_i  =  \frac{1}{1+e^{\beta_0+ \bm{\beta} \bm{X}_{i}}}.$$

The model coefficients $\bm{\beta}$ are estimated using maximum likelihood and are used to estimate $p_i$. The class of a MF detection $i$ can be predicted by thresholding the estimated $p_i$ at a particular value therefore giving a linear classification algorithm. As such, LR is a related model to Fisher's linear discriminant analysis (LDA) \citep{fisher36} and SVMs with linear kernels. Other classification methods that we considered were chosen because they take very different approaches to inducing nonlinearity, feature selection and model-fitting.

LR is an example of a feedforward neural network architecture with a single neuronal unit, single layer, and sigmoid activation function.
By adding units and layers, the neural networks extend the LR model to complex models with non-linear classification boundaries. The architectures with two or more hidden layers are generally called \textbf{deep neural networks} (DNNs) \citep{geron18}. There are many types of DNN architectures and these models that have been successfully applied in the domains of text and image analysis. In this paper we employed a multi-layer perceptron (MLP) which consists of a sequence of densly connected layers of neurons. This is the classic architecture for the data  where the feature vector does not have a hierarchical structure, as is the case for image or text data.

In this paper we considered a range of fully connected layers architectures: two hidden layers with 8 and 4 units (DNN\_8\_4), two hidden layers with 16 units each (DNN\_16\_16), two  hidden layers with 256 and 32 units (DNN\_256\_32) and three hidden layers with 13, 6 and 6 units each (DNN\_13\_6\_6). We chose {\em tanh} activations for the hidden units. The output layer for each of the networks is a single sigmoid unit and the loss function is set as the binary cross-entropy. The networks have been trained over 200 epochs with a mini-batch size of 1024 and using the ADAM optimization strategy.
We did consider deeper architectures, but they were overfitting the data  and we also considered different activation functions but there was no difference in the algorithm performance.

A \textbf{support vector machine} (SVM) is a kernel extension of a binary linear classifier  that constructs a hyperplane to separate two classes. The hyperplane is chosen so that the smallest perpendicular distance of the training data to the hyperplane (margin) is maximized. A tuning parameter (cost) controls the number of observations that are allowed to violate the margin or the hyperplane. Kernel trick is a general technique that can be applied to any optimization problem which can be rewritten so that it takes the inner products between pairs of the training observations as opposed to the observations themselves. When the inner product is replaced with a more general kernel, the observations are implicitly mapped to a higher dimensional feature space where the optimization takes place. For linear classifiers, this has the effect of fitting nonlinear decision boundaries in the original feature space. The shape of the boundary is determined by the choice of kernel and its parameterization.

\textbf{Random Forests} (RF) provide a different approach to the classification problem and to feature selection.  They grow a number of decision trees on bootstrapped samples of the training set. Each tree recursively partitions the feature space into rectangular subregions where the predicted class is the most common occurring class. At each iteration, a tree algorithm searches through all the possible split-points along a randomly selected subset of features to find a partition which minimizes the region impurity, measured by the Gini index. For a binary problem, the Gini index is given by: $2\hat{p}_m(1-\hat{p}_m)$, where $\hat{p}_m$ is the proportion of flaring observations in region $m$. A single consensus prediction is obtained from all the trees using majority vote which allows for very complex and nonlinear decision boundaries.
The total decrease in the Gini index from splitting on a feature, averaged over all trees can be taken as an estimate of that feature's importance.

\subsection{Feature selection}

For this study, we use the Marginal Relevance (MR) score to rank the features in order of their capability to discriminate between the two classes (flare/non-flare). The MR score for each feature is the ratio of the between-class to within-class sum of squares. This idea underpins many statistical methodologies and is frequently used in genetics to screen out a large number of spurious features, see, for example, \cite{dudoit02}.

 The approach to feature selection using MR screens out the unnecessary features before applying logistic regression. MR considers the information in each feature independently so the highest ranked features can be correlated and do not necessarily form the optimal subset for the purposes of classification.

For these datasets we also fitted Lasso \citep{tibshirani96}, a model related to LR, but where the coefficients $\beta$ are simultaneously shrunk to zero using a penalty which is controlled by a tuning parameter. Lasso provides a more sophisticated approach to feature selection than MR and simultaneously reduces dimensionality of the feature space and performs classification. Lasso is implemented in R package glmnet \citep{friedman09}.

All features were used to train the  nonlinear classifiers. SVMs combine all the feature information into a distance matrix (kernel) and can cope with correlated inputs and a small number of spurious features. In RF all the features are used to grow the trees and at each iteration randomly selected subsets are jointly considered for subdividing the feature space.

\subsection{Cross-validation}\label{sec:CV}

For consistency and comparison with \cite{ahmed11} we use the MF detections from April 1996 - December 2000 and January 2003 - December 2008 to train the classification algorithms and the MF detections from January 2001 - December 2002 and January 2009 to December 2010 comprise the test set.

The number of flaring/non-flaring SMART detections in the training and testing sets for both the full and NOAA AR data are shown in Table~\ref{table:detections}.

The training set is further subsampled by randomly drawing 200 instances of flares and 200 instances of non-flares to form 50 smaller training sets.

The full SMART dataset contains 490,997 non-flaring and 27,244 (5.4\%) flaring instances and therefore exhibits a large class imbalance. In the NOAA AR dataset 18.6\% of detections were classed as flares. Class imbalance is a common problem and has received a great deal of attention in classification literature, see, for example \cite{chawla04}.  In construction of the subsampled training sets we uniformly sampled instances of flares and non-flares but adjusted the mixture of the classes, an approach known as case-control sampling. Logistic regression models fitted to subsamples can be converted to a valid model using a simple adjustment to the intercept, see \cite{fithian14}.

 \begin{table}[ht]
\caption{The number of flaring/non-flaring SMART detections in the full and reduced datasets.}\label{table:detections}
\begin{tabular}{lllll}
  \hline
 & \multicolumn{2}{l}{Full SMART dataset}& \multicolumn{2}{l}{NOAA AR  dataset}\\
\hline
  & Training set & Testing  set & Training set & Testing  set \\
     \hline
flare & 16673  &10571 &1137 &707 \\
non-flare& 313617  & 177380&5272 & 2789 \\
   \hline
\end{tabular}
\end{table}

\subsection{Forecast performance measures}\label{sec:forecast}

In a binary classification problem we can designate one outcome as positive (flare) and the other as negative (no flare). For algorithms with probabilistic outputs, binary forecasts are obtained by thresholding $p$, e.g. predicting a flare if the estimated  $p>0.5$. A confusion matrix is constructed by cross-tabulating the predicted with the observed classes. This presents the number of true positives TP (flare predicted and observed), false positives FP (flare predicted but not observed), true negatives TN (no flare predicted and none observed) and false negatives FN (no flare predicted but observed).

The true positive rate (TPR), or sensitivity, is the proportion of correctly classified flares out of all the flares observed in the sample TPR = TP/ (TP+FN).  The true negative rate (TNR), or specificity, is the proportion of true negatives out of all the non-flaring instances.
The false positive rate is FPR =  1 - TNR and the false negative rate is FNR = 1 - TPR.

A classifier that performs well will give a high TPR and TNR and, consequently, low FPR and FNR. For classifiers that give probabilistic outputs, sensitivity (TPR) can be increased by lowering the threshold of $p$, but this automatically increases the FPR. An optimal choice of the threshold is context dependent: the cost of FNs might be higher than FPs.
For a ($0,1$) range of thresholds, receiver operating characteristic (ROC) curve plots the TPR vs FPR. ROC curve and the corresponding area under the ROC curve (AUC) are used for comparing the performance of algorithms over the entire range of thresholds. The ideal ROC curve is in the top left corner, giving high TPR and a low FPR, and the maximum possible value for AUC is 1.

For a single threshold or for classifiers with non-probabilistic outputs, the elements of the confusion matrix can be combined in a number of ways to obtain a single measure of the performance of a given method.

Accuracy (ACC) gives the proportion of correctly classified observations over both classes.

True skill statistic (TSS), \citep{youden50, hanssen65} combines the sensitivity and specificity by taking TSS = TPR + TNR - 1.

Heidke skill score (HSS) \citep{heidke26} measures the fraction of correct predictions after adjusting for the predictions which would be correct due to random chance.

For more details on forecast performance measures used in solar flare literature see \cite{bloomfield12, barnesleka08, allClear16}.

\section{Results}\label{sec:results}

LR with top three features selected by MR (LR) and SVMs were trained on 50 subsampled training sets. The results for the SVM were obtained from R package e1071 \citep{e1071}, with a Gaussian kernel with the bandwidth parameter set to $0.03$ for the full SMART dataset and $0.01$ for the NOAA AR dataset. The cost of constraints violation was set to $1$.

LR with top three features selected by MR (LR3), the full set of features (LR13), RF and DNN were trained on the full training set. The DNN architectures were: two layers with 8 and 4 units (DNN\_8\_4), two layers with 16 units each (DNN\_16\_16), two layers, 256 and 32 units (DNN\_256\_32) and three layers 13,6 and 6 units each (DNN\_13\_6\_6).

For RF, $500$ trees were grown, where at each iteration three variables were randomly sampled as candidates for each split. The tuning parameters of RF and SVMs were tuned over grids using cross-validation of the training set. For support vector machines (SVMs) we tried out Gaussian, Anova and Laplacian kernels and Bayesian Kernel Projection Classifier (BKPC) \citep{domijan09}, a sparse Bayesian variant using lower dimensional projections of the data in the feature space. All three kernels and BKPC performed equally well in the training set at the optimal values of their kernel parameters so, in the paper we present the results of the Gaussian kernel as it is best known.
RF is implemented in R library randomForest \citep{liaw02}. MR algorithm is implemented in R library BKPC \citep{domijan16}. DNNs were fitted using Keras library in R \citep{keras17}.

For the purposes of analysis, some of the features were log transformed (high-gradient neutral-line length in the region, neutral-line length in the region,  Falconer's $WL_{SG}$ value, Schrijver's $R$ value, total un-signed magnetic flux and flux emergence rate). Same transformations were found to be adequate for both the full SMART dataset and the NOAA AR dataset. For both datasets, the features in the training sets were scaled to have zero mean and unit variance and the same scaling was then applied to the test sets.

The forecast performance measures (TPR, TNR, TSS,  ACC and HSS), described in Section~\ref{sec:forecast},  were calculated for the test set at a range of thresholds of $p$.

For the classifiers trained on 50 subsampled training sets,  50 classification rules were obtained and consequently the median values, $2.5$th and $97.5$th percentiles of the resulting forecast performance measures are reported. This can be used to assess the sensitivity of the algorithms to the choice of the training sets.

\subsection{Full SMART dataset}\label{sec:resultsFULL}

\begin{figure}[t!]
\centerline{
\hspace*{0.015\textwidth}
  \includegraphics[width=7.5cm, height=5cm]{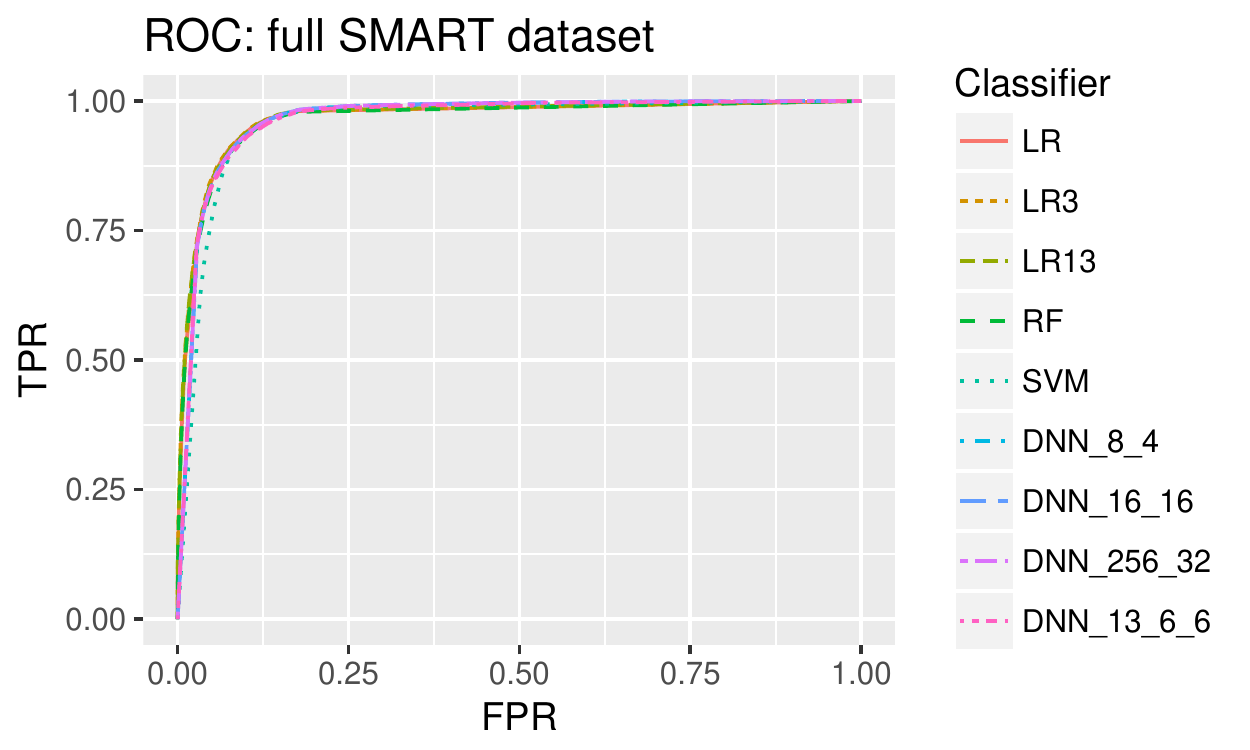}
  \hspace*{-0.03\textwidth}
    \includegraphics[width=7.5cm, height=5cm]{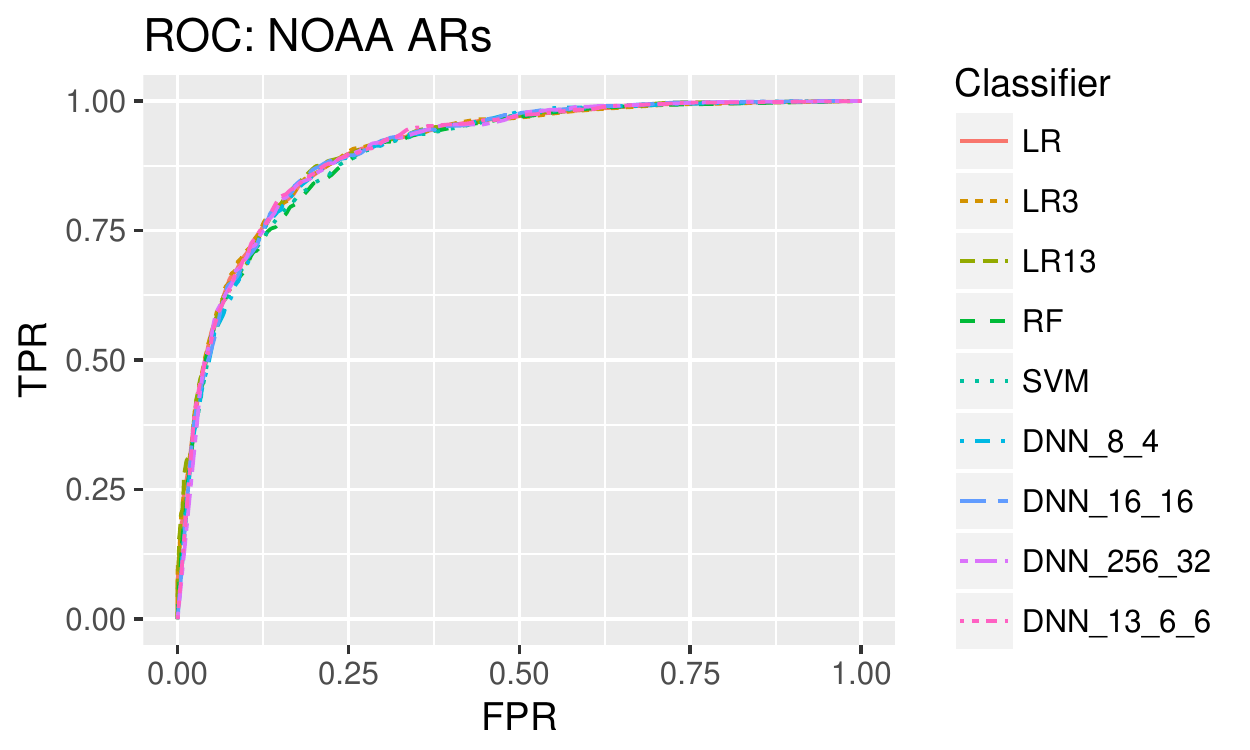}
    }
     %
       \vspace{-0.42\textwidth}   
     \centerline{\Large \bf     
      \hspace{-0.088 \textwidth} \small{(a)}
      \hspace{0.545\textwidth}  \small{(b)}
         \hfill}
     \vspace{0.38\textwidth}    
   \caption{ROC curves, calculated for the training set for LR (LR with 3 features trained on subsamples), LR3 (LR with 3 features trained on the full training set),  LR13 (LR with all features trained on the full training set), RF, SVM and the DNN architectures. (a) Full SMART dataset, (b) NOAA ARs dataset. }\label{fig:roccompare}
\end{figure}

Figure~\ref{fig:roccompare}(a) shows the ROC curves plotted for all the classifiers for the full SMART dataset. For the algorithms trained on 50 subsampled training sets (LR and SVM), the ROC curve is obtained from the median TPR and FPR over the $(0,1)$ range of thresholds. The ROC curves are very close and show that after careful tuning, all models perform equally well and converge to the same results in terms of the performance measures.

\begin{table}[t]
\caption{AUC, the highest TSS and HSS for all the classifiers. The first two (LR and SVM) were fitted to subsampled training sets. The rest were trained on the full training set. }\label{table:TopTSS}
\begin{tabular}{llll}
  \hline
 Classifier & TSS& HSS & AUC\\
  \hline

LR & 0.84 (0.83, 0.84)& 0.63 (0.60, 0.64)& 0.966 (0.962, 0.968)\\
SVM & 0.83 (0.83, 0.84)& 0.56 (0.51, 0.60)& 0.949 (0.942, 0.956) \\
LR3 & 0.84 & 0.64& 0.967\\
LR13 & 0.84  & 0.64&0.967\\
RF & 0.83  & 0.63&0.964\\
DNN\_8\_4 & 0.83 & 0.63& 0.966\\
DNN\_16\_16 & 0.83  & 0.63&0.966\\
DNN\_256\_32 & 0.83  & 0.63&0.966\\
DNN\_13\_6\_6 & 0.83  & 0.64&0.965\\

   \hline
\end{tabular}
\end{table}

AUC, the highest TSS and HSS for all classifiers for the full SMART data are given in Table~\ref{table:TopTSS}. For algorithms trained on 50 subsamples, the reported value is the median with $2.5$th and $97.5$th percentiles given in brackets. AUC ranged from 0.949 to 0.967, TSS ranged from 0.83 to 0.84 and HSS ranged from 0.56 to 0.64. This again shows that no model convincingly outperformed the others in terms of predictive ability for this dataset. The linear model with only three features works as well as the more complex models that allow for nonlinear classification boundaries. Likewise, including extra features in the linear model did not improve performance. The results for LR with three features are the same for the algorithm trained on the subsampled training sets and the full training set (LR and LR3), showing that small datasets of 400 instances are sufficient to train this model. Narrow confidence bands for LR and SVM indicate that the classification results are consistent across the subsampled training sets.

For the logistic regression algorithm with the three input features trained on the subsampled training sets (LR) the median values for TSS, TPR, TNR, ACC and HSS at each threshold are presented in Table~\ref{table:test} in the Appendix. The 2.5th and 97.5th percentiles for TSS and HSS are given in brackets. For comparison, Table~\ref{table:testLRfull} presents the results for the same algorithm trained on the full training set (LR3).

For LR, the highest TSS of 0.84 is obtained at the thresholds between 0.04 to 0.08 which give the TPR in the range of  0.96 and 0.92 and TNR of 0.88 and 0.91 respectively. The results from \cite{ahmed11} give a TPR of 0.523, TNR of 0.989 with HSS of 0.595 for the machine learning algorithm and TPR of 0.814 and HSS of 0.512 (TNR is not reported) for the automated solar activity prediction (ASAP).
Area under ROC curve (not reported by \cite{ahmed11}) was calculated for the 50 curves and the median AUC value for LR is 0.966 with 2.5th and 97.5th percentiles of 0.962, 0.968. For the same model trained on the full training set using three features (LR3), the AUC was 0.967. Using all features in the model (LR13) did not increase AUC from 0.967.

\subsection{Choice of skill scores and threshold}

The LR model fits a sigmoid surface over the range of $\bm{X}$ and the decision boundary separating the two predicted classes at any threshold is linear.

Note that in this dataset a single MF is tracked though time and will be recorded multiple times throughout its lifetime.  For the purpose of this analysis, all MF detections are treated as individual measurements. This can pose a problem for interpreting the probabilistic inference of LR which is underpinned by the independence assumption: the standard error estimates for the coefficients are no longer reliable and $p$ cannot be interpreted as probability. However, in this paper, we do not make use of probabilistic inference and we treat the logistic regression model as a deterministic linear classifier, where the choice of thresholding value of estimated $p$ is based on context requirements: comparing the acceptable levels of true positive and true negative rates for different thresholds.  Furthermore, by training algorithms on very small subsets of the original data (100 instances of flares randomly drawn from 16,673 and 300 non-flares from 313,617) one is unlikely to get many detections of the same MF in the same sub-sample, which helps evade this problem. Alternatively, one could enforce the randomly drawn detections to have large enough time cadences between them (e.g. more than two weeks) in order to ensure that the same MF is not recorded in the same training set multiple times throughout its lifetime.

\begin{figure}[t!]
\centerline{
  \includegraphics[width=10cm, height=7cm]{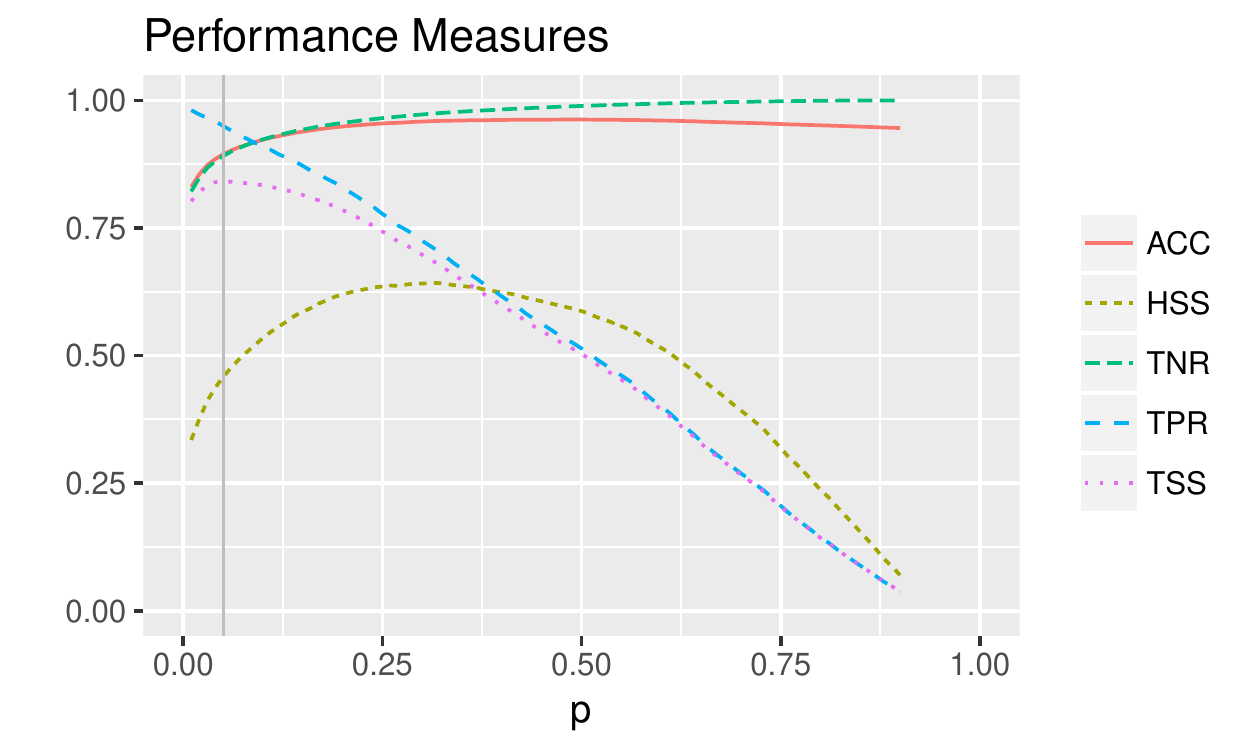}
  }
   \caption{Sensitivity (TPR), specificity (TNR), accuracy, TSS and HSS over a range of probability thresholds. Estimated probability was obtained from the logistic regression with three features on the full SMART dataset. The proportion of flares in the training dataset is 0.05 (vertical line). }\label{fig:SENSSPECfull}
\end{figure}

Figure~\ref{fig:SENSSPECfull} shows the calculated sensitivity (TPR), specificity (TNR), ACC, TSS and HSS over a range of probability thresholds (0.01 to 0.99 in steps of 0.01), where $\hat{p}$ was estimated from the LR3 model.  The proportion of flaring instances in the full training set is $0.05$, shown as the vertical line on the graph. This figure shows increase in TNR  at the expense of TPR with increase of the threshold. At the lowest threshold $p = 0.01$, the $82\%$ of non-flares are correctly identified and this increases to $89\%$ at 0.05. Likewise, $98\%$ of flares are correctly predicted at the threshold of 0.01 and $95\%$ are correctly identified at the threshold of 0.05. The maximum estimated TSS = TNR + TPR - 1 is 0.84. Accuracy, a measure of overall error rate is maximized at the threshold of $p=0.5$, which gives TNR of $99\%$, but misclassifies $50\%$ of the flares.
This illustrates how ACC is a very poor choice of metric for data with a large class imbalance. At threshold of 1, classifying all observations as negative, one will still get a $95\%$ accuracy score and choosing a threshold of $p=0.5$ will misclassify over half the flare detections. Given that the assumptions for probabilistic inference in LR are met, the $p$ estimates the likelihood that an observation is going to flare given its feature information. Before fitting the model and utilizing the feature information, the probability that a randomly selected detection will flare is 0.05. Unlike LDA, the prior information about flare prevalence is not incorporated in the model. Thus it is sensible to take this prior as a threshold as opposed to $0.5$. For this threshold, the algorithm will classify a detection as flare if the estimated $p$ is greater than the probability we would assign to a randomly selected detection. Figure~\ref{fig:SENSSPECfull} shows that this threshold strikes a sensible balance between TPR and TNR and indicates that TSS is a better choice than HSS for imbalanced data. HSS is maximized at a higher threshold of $0.3$ with TPR of 0.72.

ROC curves allow for comparison of classifiers with probabilistic outputs over the range of thresholds. AUC summarises the forecast performance in a single score, however, one could argue that comparing algorithm performance at an `optimal' threshold is more useful than over the entire range. When comparing classifier performance using skill scores, we argue that it is useful to plot skill score curves over the threshold range.

\begin{figure}[t!]
\centerline{
\hspace*{0.015\textwidth}
  \includegraphics[width=7.5cm, height=5cm]{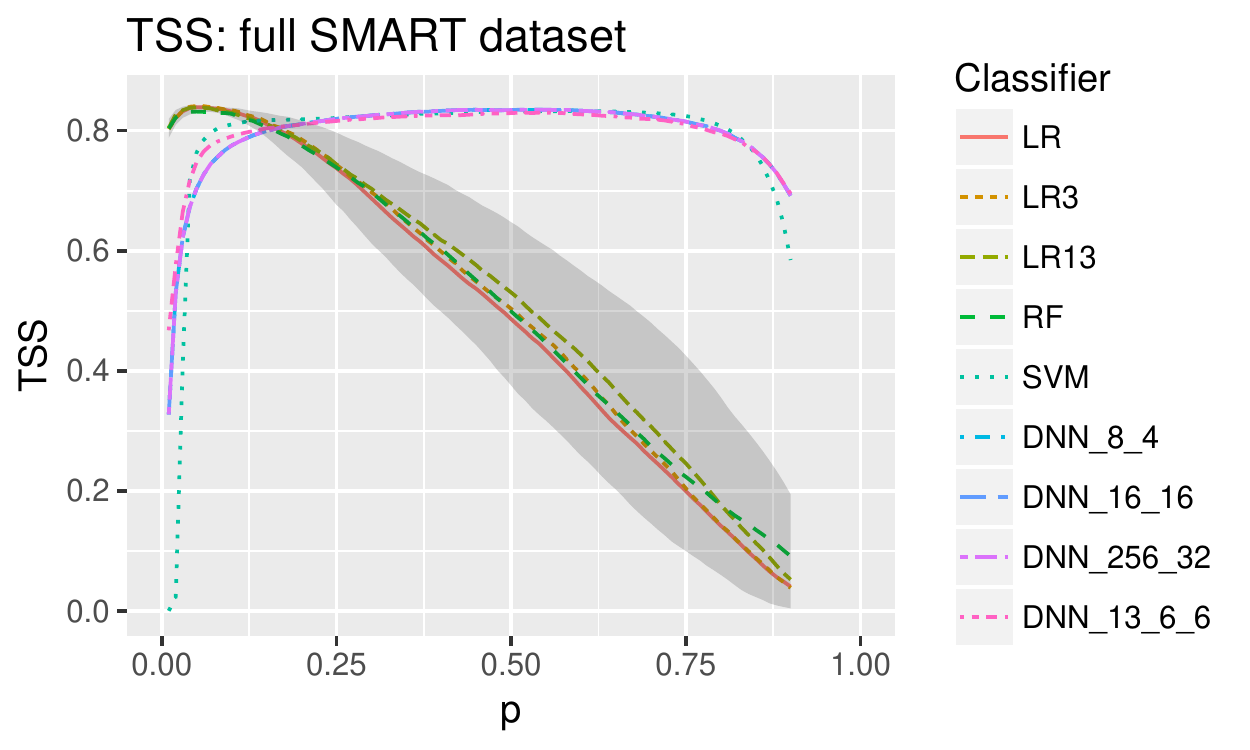}
  \hspace*{-0.03\textwidth}
  \includegraphics[width=7.5cm, height=5cm]{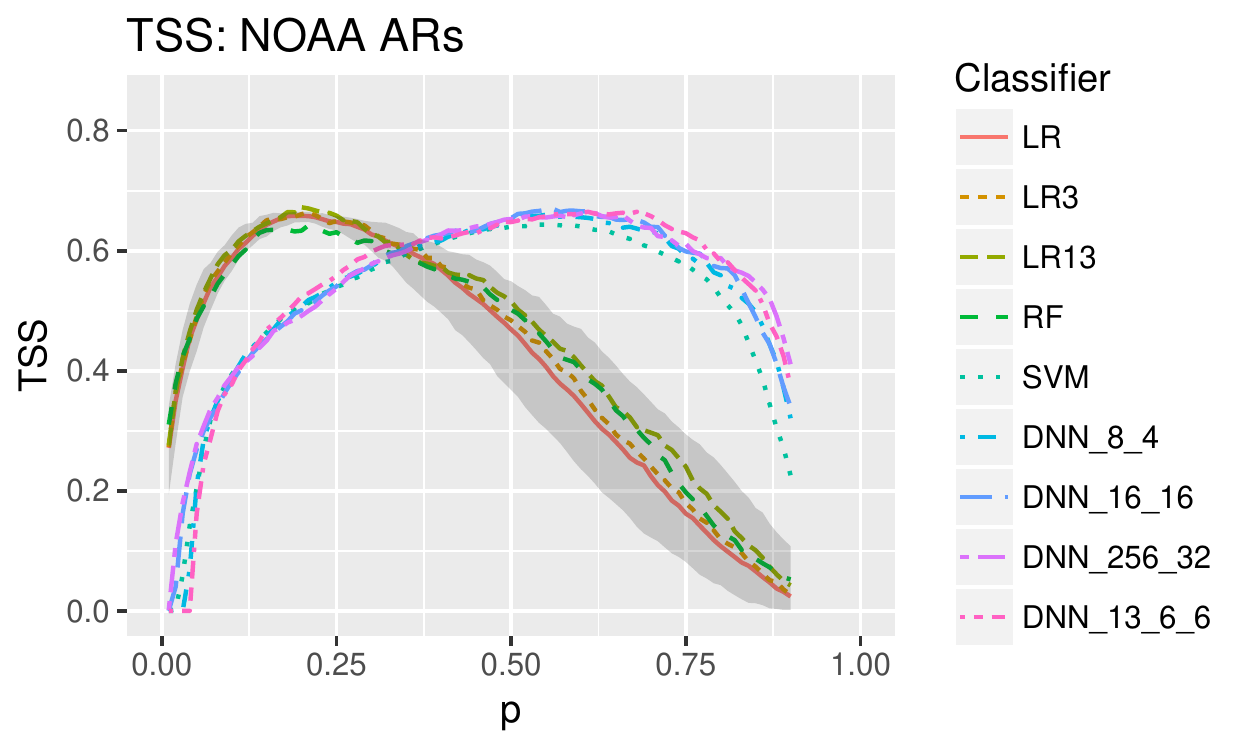}
}
       \vspace{-0.42\textwidth}   
     \centerline{\Large \bf     
      \hspace{-0.095 \textwidth} \small{(a)}
      \hspace{0.545\textwidth}  \small{(b)}
         \hfill}
     \vspace{0.38\textwidth}    
   \caption{TSS or median TSS curve for all the classifiers.  LR (LR with 3 features trained on subsamples), LR3 (LR with 3 features trained on the full training set),  LR13 (LR with all features trained on the full training set), RF, SVM and four DNN architectures. TSS curve for LR (full line) has $2.5$th and $97.5$th percentile band. (a) Full SMART dataset, (b) NOAA AR dataset.}\label{fig:tsscompare}
\end{figure}

Figure~\ref{fig:tsscompare} shows the TSS curves for different classifiers (for LR the curve is median TSS with $2.5$th and $97.5$th percentile band). Figure~\ref{fig:tsscompare}(a) shows the TSS curves from the algorithms trained and tested on the full SMART dataset. The maximum TSS values obtained from all classifiers are very close (ranging from 0.82-0.84) but are obtained at different thresholds for $p$ since the TSS curves differ in shape. This is due to the fact that the probability of flare is estimated differently in these models. For the RF the estimate of $p$ is non-parametric. By default SVM produces categorical outputs, however probabilistic extensions exist and the R implementation fits a probabilistic regression model that assumes (zero-mean) Laplace-distributed errors for the predictions. For DNNs, in order to balance the classes, the class weight of 20 was used for the flare class labels in the computation of the loss function, which is equivalent to upsampling to match the majority class. Therefore, for the DNN models, TSS is optimised at a threshold of 0.5.

\subsection{NOAA AR dataset}\label{sec:resultsRED}

Figure~\ref{fig:roccompare}(b) shows the ROC curves plotted for all the classifiers for the NOAA AR data. TSS curves are given in Figure~\ref{fig:tsscompare}(b). For LR the curve is median TSS with $2.5$th and $97.5$th percentile band. Compared  to the results of the algorithms trained and tested on the full SMART data, the performance of all algorithms is significantly weaker if trained and tested on the NOAA AR dataset, however this is still competitive with the results reported elsewhere in the solar flare forecasting literature, for example see \cite{allClear16}. 

\begin{table}[t]
\caption{NOAA AR data: AUC, the highest TSS and HSS for all the classifiers. The first two (LR and SVM) were fitted to subsampled training sets. The rest were trained on the full training set. }\label{table:TopTSSRD}
\begin{tabular}{llll}
  \hline
 Classifier & TSS& HSS & AUC\\
  \hline
LR & 0.66 (0.658, 0.664)& 0.59 (0.57, 0.59)& 0.90 (0.90, 0.90)\\
SVM & 0.64  (0.63, 0.66)& 0.57 (0.55, 0.58)& 0.90 (0.89, 0.90) \\
  LR3 & 0.66 & 0.59& 0.91\\
  LR13 & 0.67  & 0.58& 0.91\\
  RF & 0.64  & 0.57&0.90\\
  DNN\_8\_4 & 0.66 & 0.57& 0.90\\
  DNN\_16\_16 & 0.67  & 0.59&0.90\\
  DNN\_256\_32 & 0.66  & 0.59&0.90\\
  DNN\_13\_6\_6 & 0.66  & 0.58&0.90\\
   \hline
\end{tabular}
\end{table}

For the NOAA AR dataset, AUC, the highest TSS and HSS for all classifiers are given in Table~\ref{table:TopTSSRD}. AUC ranged from 0.9 to 0.91, TSS ranged from 0.64 to 0.67 and HSS ranged from 0.57 to 0.59. For logistic regression algorithms (LR, LR3, LR13) and RF, the TSS is optimised at the threshold of $p \approx 0.18$ which is the prevalence of flares in the training set of this data. For DNN architectures the class weight of 5 was used for the flare class labels in the computation of the loss function. For full output of LR and LR13 results, see Table~\ref{table:testNOAA} and Table~\ref{table:testDNNfull} in the Appendix).

The results show that the algorithms with top three features (LR and LR3) perform as well as the linear model with all features (LR13) and all the nonlinear algorithms (DNNs, RF and SVM). In addition, small datasets of 400 instances are sufficient to train the linear model. Narrow confidence bands for LR and SVM indicate that the classification results are consistent across the subsampled training sets.

\subsection{Feature analysis and selection}

The marginal relevance score for each feature was derived from the data used to train the classification algorithm (detections recorded from April 1996 - December 2000 and January 2003 - December 2008).  Features in order of their marginal relevance derived from the full and the NOAA AR dataset are given in Table~\ref{table:features}. The third and fourth columns give the importance order of the top six features obtained from the Random Forest in  both datasets. The top features selected by Lasso are given in columns five and six. Column titles with (R) denotes the importance measures were derived for the reduced NOAA AR dataset.

MR selects high-gradient neutral-line length in the region (LsgMm), maximum gradient along polarity inversion line (MxGradGpMm) and neutral-line length in the region (LnlMm) as the top three features for both full SMART and NOAA AR dataset. For both datasets, the best performing and most parsimonious Lasso model had three features, but selected area of the region  (AreaMmsq) or total un-signed magnetic flux (BfluxMx) instead of maximum gradient along PIL (MxGradGpMm). These algorithms had the same classification performance as LR3 and LR13. In addition to neutral-line length in the region (LnlMm), features with highest importance selected by RF were Schrijver's $R$ value (RvalMx), total un-signed magnetic flux (BfluxMx) and Falconer's $WL_{SG}$ value (WLsgGpMm).

The forecast performance measures for the Lasso models, RFs and LR with three or more features were similar in both datasets. Many other approaches to feature selection exist, but this indicates that there is little scope for improvement with more thorough exploration of the feature space.

\begin{sidewaystable}[ht]
\caption{SMART magnetic features in order of their marginal relevance obtained from the training dataset for the dataset of all SMART detections. 
The second column MR(R) is the MR feature ranking obtained from the training set of the NOAA AR  detections only. Columns RF and RF(R) present  the variable importance order obtained from running the Random Forest on the training set of the full and NOAA AR  (R) datasets. Columns Lasso and Lasso (R) present the features selected in the four sparsest models fitted to the training set of the full and NOAA AR  (R) datasets.}\label{table:features}

\begin{center}

 \small{
\begin{tabular}{rrrrrrll}
  \hline
MR & MR(R) & RF& RF(R)&Lasso&Lasso(R)& Feature &  Description\\
  \hline
1 &1  &5  &5  & 1 &1& LsgMm &High-gradient neutral-line length in the region\\
2 &3  &$-$&6  & $-$&$-$& MxGradGpMm &  Maximum gradient along polarity inversion line \\
3 &2  &1  &1 & 1 &1& LnlMm &  Neutral-line length in the region\\
4 &5  &4  &3 &$-$&$-$& WLsgGpMm &  Falconer's $WL_{SG}$ value\\
5 &6  &6  &$-$&1& $-$&AreaMmsq & Area of the region\\
6 &7  &2  &2&$-$&$-$& RvalMx &  Schrijver's $R$ value\\
7 &9  &$-$&$-$&$-$&$-$& B &   Largest magnetic field value \\
8 &8  &$-$&$-$&$-$& $-$&HGlonwdth &  Heliographic longitudinal extent\\
9 &4  &3  &4&$-$& 1&BfluxMx &  Total un-signed magnetic flux\\
10&10 &$-$&$-$&$-$&$-$& HGlatwdth & Heliographic latitudinal extent\\
11&11 &$-$&$-$&$-$& $-$&MednGrad & Median gradient along the neutral line \\
12&12 &$-$&$-$ &$-$& $-$&Bfluximb &  Flux imbalance fraction in the region\\
13&13 &$-$&$-$&$-$& $-$&DBfluxDtMx & Flux emergence rate\\
   \hline
\end{tabular}
 }

\end{center}
\end{sidewaystable}

Figure~\ref{fig:topfeats} and \ref{fig:topfeats2} in the Appendix plot the marginal densities from the training set of the full SMART dataset and NOAA AR dataset of some of the top features selected by MR, Lasso  and  RF. Large peaks close to 0 in the non-flare distribution show that the full SMART dataset is dominated by the small magnetic flux regions that do possess many of the properties that SMART calculates and never flare.
All the displayed features  contain information on whether an observation is likely to be a flare, however, all contain a significant amount of overlap in the distributions of flaring and non-flaring regions. Schrijver's $R$ and Falconer's $WL_{SG}$ values show a clear separation between two clusters, one of which has a very low proportion of flares, whereas the density plots for most other features indicate a steady increase in the proportion of flares, which gives an insight into why splitting on these variables would lead to a decrease in the Gini index.

\begin{figure}[t!]
\centerline{
\hspace*{0.015\textwidth}
  \includegraphics[width=8cm,height=8cm]{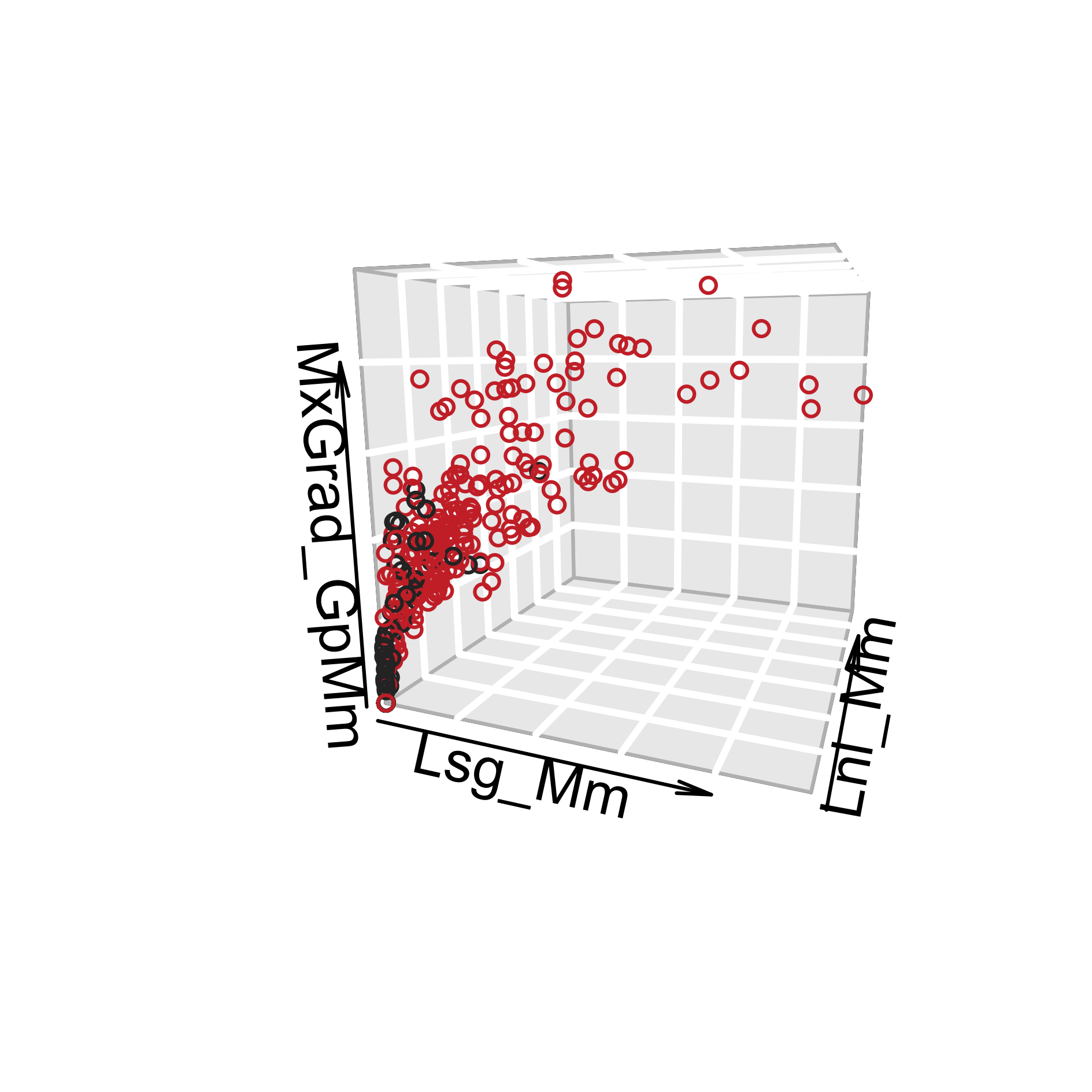}
  \hspace*{-0.03\textwidth}
  \includegraphics[width=8cm,height=8cm]{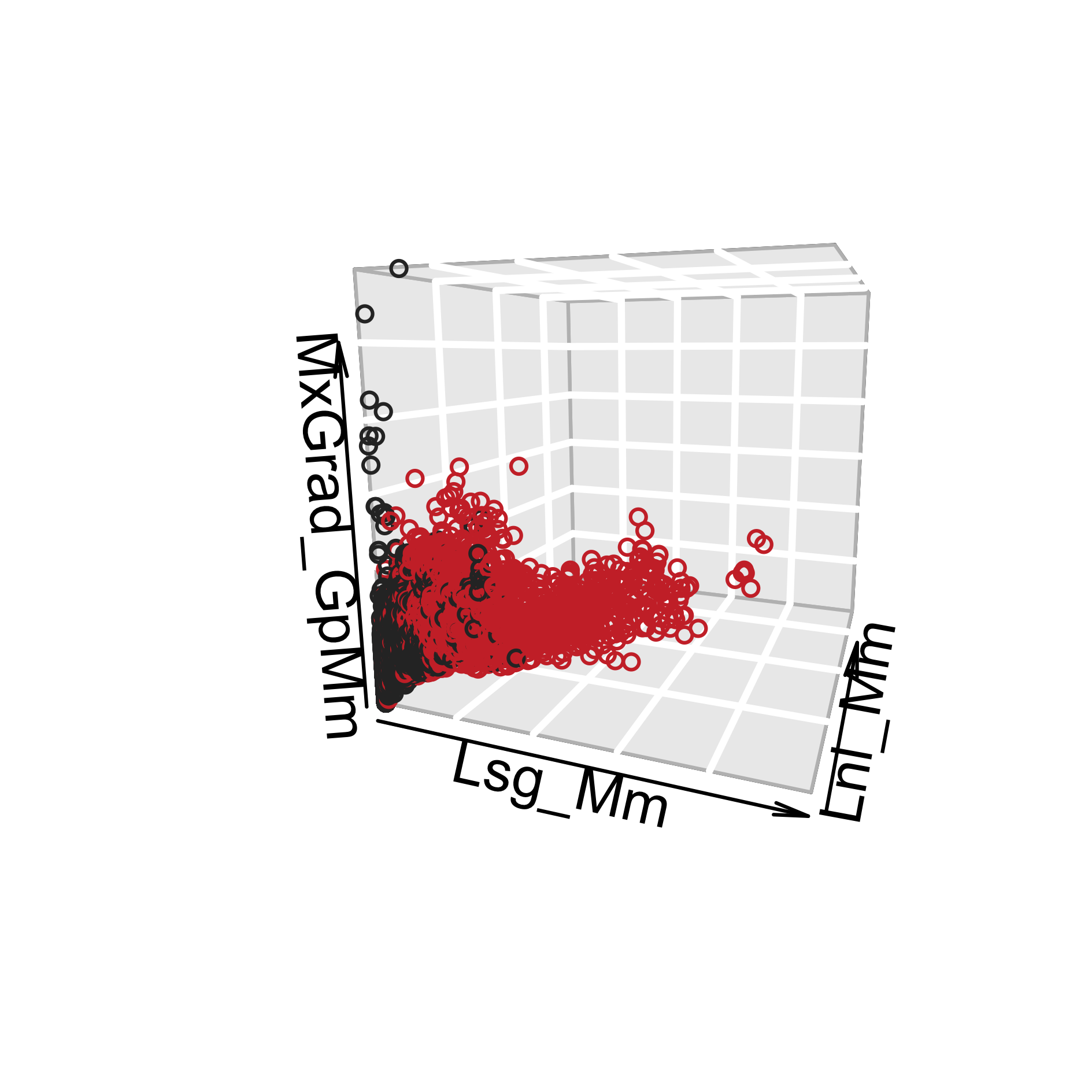}
  }
  \caption{MF detections in the (a) one training dataset and (b)  testing dataset, in 3 dimensions (3 features with the highest marginal relevance), red = flare, black = no flare.}\label{fig:train3d}
\end{figure}

Figure~\ref{fig:train3d} shows one subsampled training set and test set of the full dataset of all SMART detections in three dimensions corresponding to features with the highest marginal relevance. The detections are colored by their class (flare/non-flare). In the test set, a large proportion of detections is located around 0, but due to over-plotting, it is harder to see how these observations dominate the dataset compared to density plots. These plots show that whereas the features contain information about classes, there in an overlap between them in this feature space - the classes are not perfectly separable. The shape of the classification boundary will not change this and the scope for improvement when using more complex algorithms in this feature space is limited.

\section{Discussion}\label{sec:discussion}

In order to classify MF detections we used a number of classification approaches including binary logistic regression, SVMs, RFs and a set of DNN architectures.
Categorical forecasts were obtained by thresholding the estimated probability from these models. Skill scores, curves of TSS, ROC curves and area under the ROC were used to compare the performance of the classification approaches. We discussed the choice of skill scores and optimal thresholds for various model settings.

The flare prediction results that we obtained from the linear classifier with a very sparse subset of features compare favorably to those found elsewhere in the literature and show a significant improvement on the results of the previous analysis of this data. We found that, in terms of classification performance, there was no benefit in using more features or more flexible models that allow for nonlinear classification boundaries as all approaches converged to the same result. Furthermore, we found no decrease in performance when training the algorithms on very small subsampled training sets.

By plotting the data in the selected dimensions we see that the classes are not perfectly separable in the space of SMART features and that there is limited scope for improvement in using more complex algorithms on this dataset.

A better performance, however, might be obtained by using the deep learning networks to learn the forecasting patterns directly from magnetograms of solar active regions as opposed to using the features computed from the magnetograms.  Some work on DNNs for solar flare prediction has been done by \cite{nishizuka18} and \cite{huang18}.

Direct comparisons with other published methods are difficult because of differences in data sets, the definition of an event, and evaluation and reporting of classification results \cite{allClear16}. It would be of interest to carry out a comparative study of classification algorithms, such as presented here, to the Space-weather HMI Active Region Patch (SHARP) \citep{bobra2014} data from the \emph{Solar Dynamics Observatory's}(SDO) Helioseismic and Magnetic Imager (HMI). The data have previously been analysed by \cite{bobra2015} and \cite{liudeng17} who used SVMs and Random Forests.

The work presented in this paper is fully reproducible with code for variable selection, sub-sampling and classification available via GitHub.

\appendix

\begin{figure}[t!]
\centerline{
\hspace*{0.015\textwidth}
\includegraphics[width=7cm,height=5cm]{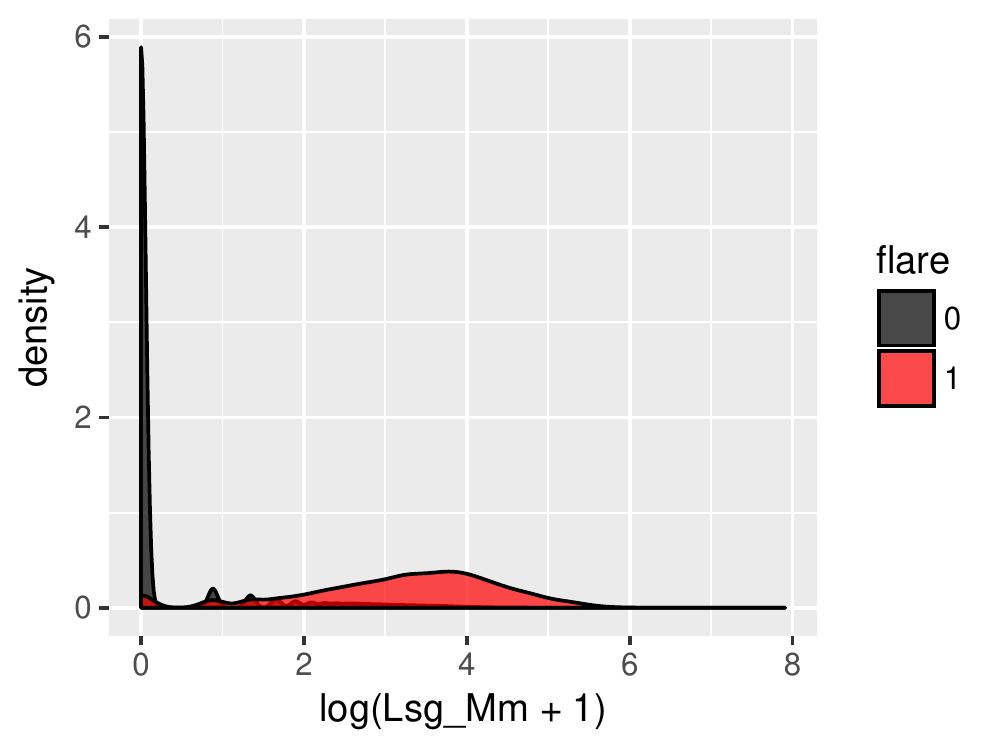}
\hspace*{-0.03\textwidth}
\includegraphics[width=7cm,height=5cm]{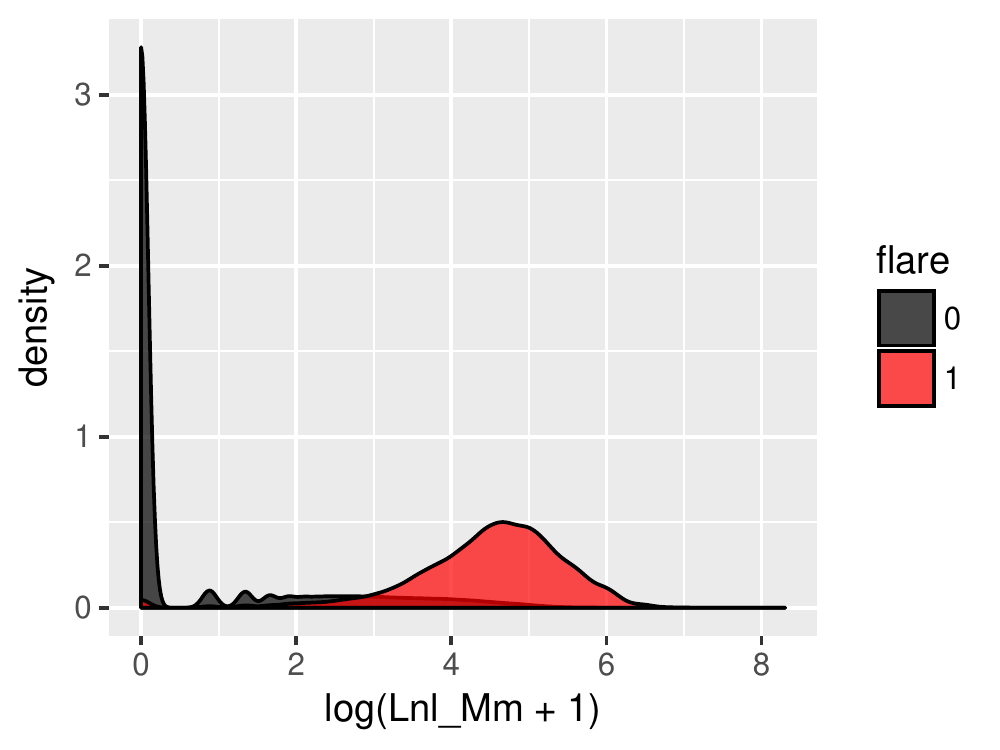}
}
\centerline{
\hspace*{0.015\textwidth}
\includegraphics[width=7cm,height=5cm]{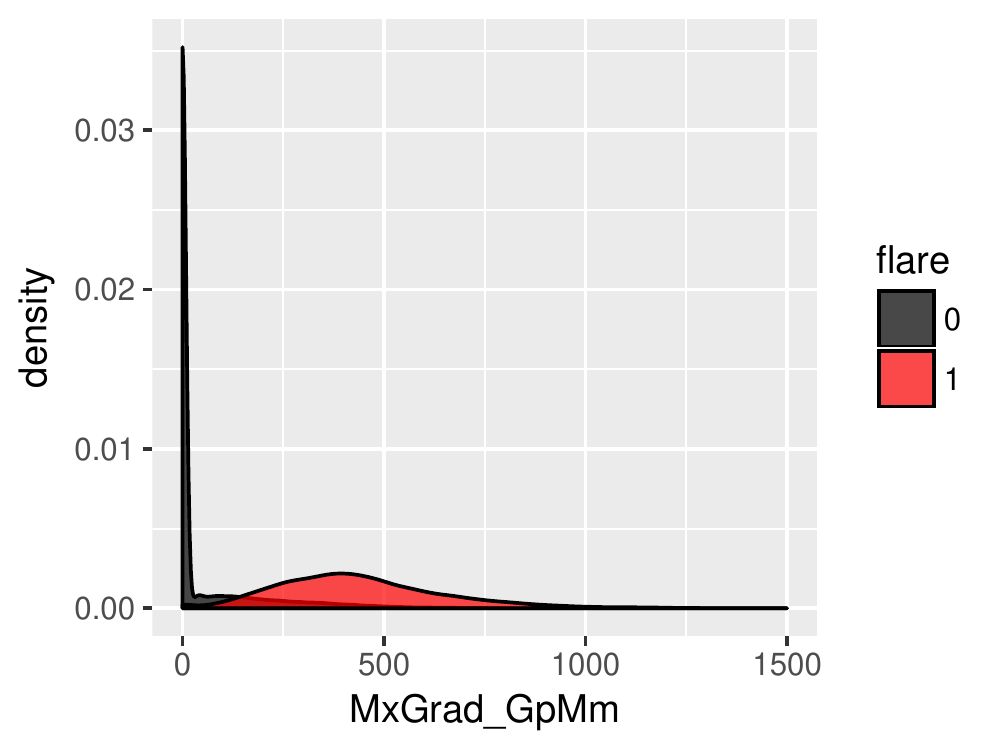}
\hspace*{-0.03\textwidth}
\includegraphics[width=7cm,height=5cm]{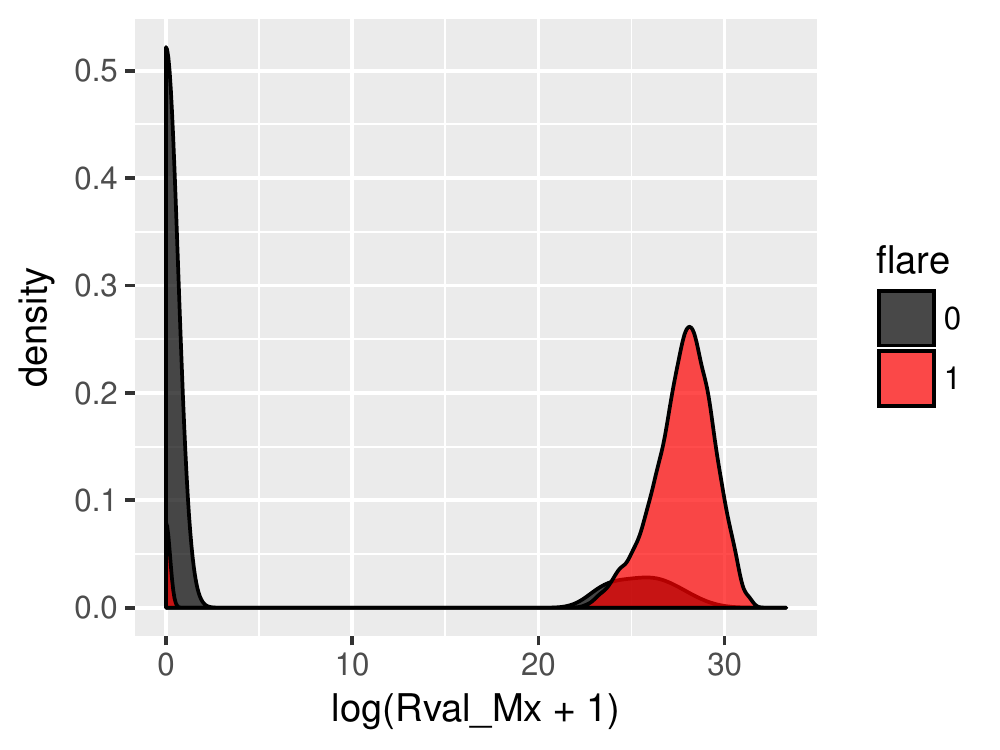}
}
\centerline{
\hspace*{0.015\textwidth}
\includegraphics[width=7cm,height=5cm]{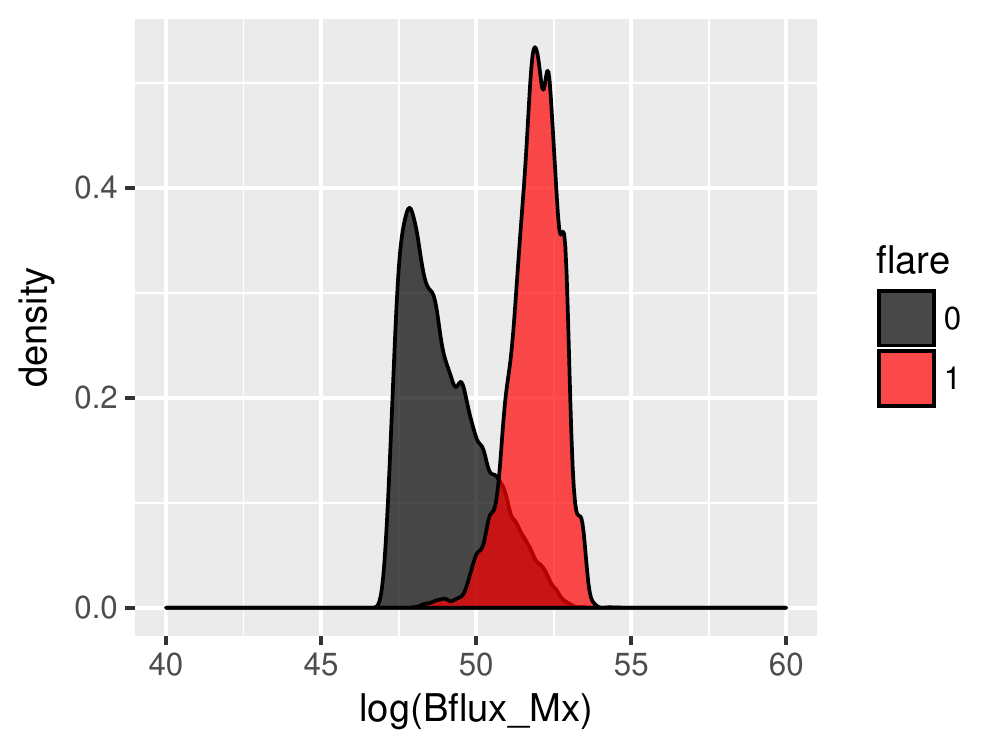}
\hspace*{-0.03\textwidth}
\includegraphics[width=7cm,height=5cm]{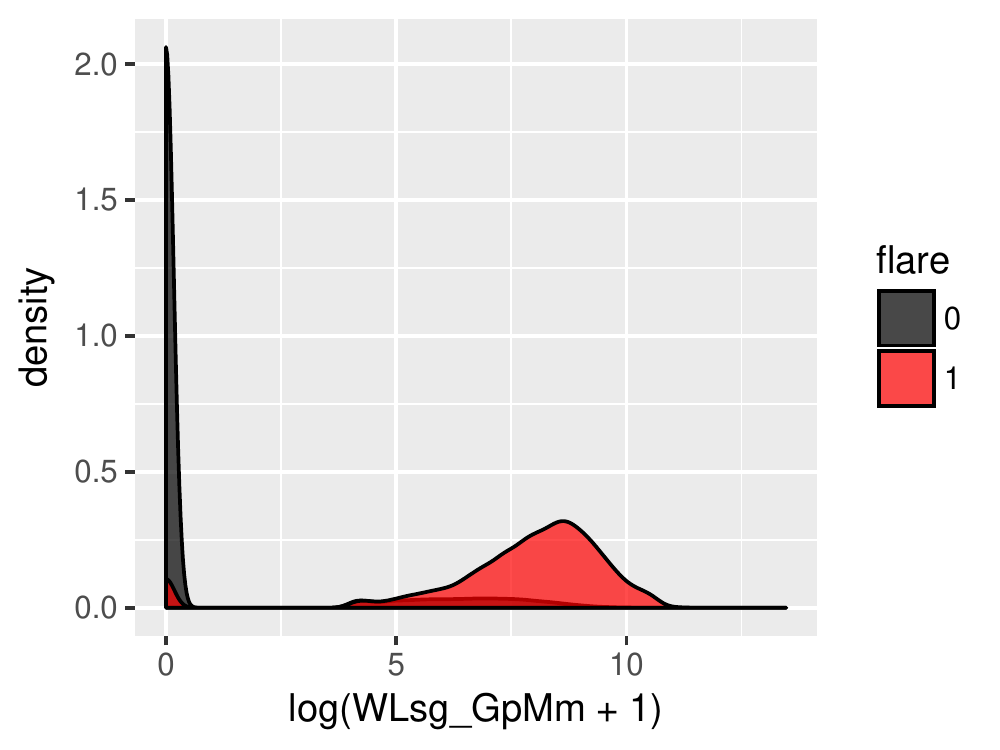}
}
  \caption{The marginal densities of the top features in the full SMART dataset selected by MR, Lasso and RF algorithms.}\label{fig:topfeats}
\end{figure}

\begin{figure}[t!]
\centerline{
\hspace*{0.015\textwidth}
\includegraphics[width=7cm,height=5cm]{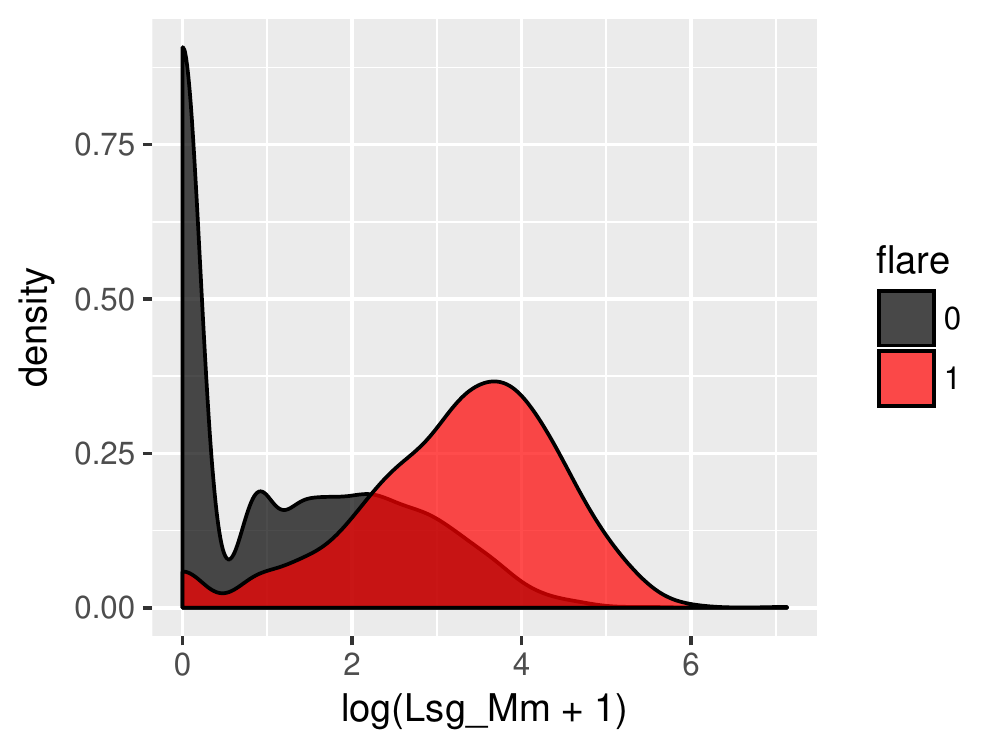}
\hspace*{-0.03\textwidth}
\includegraphics[width=7cm,height=5cm]{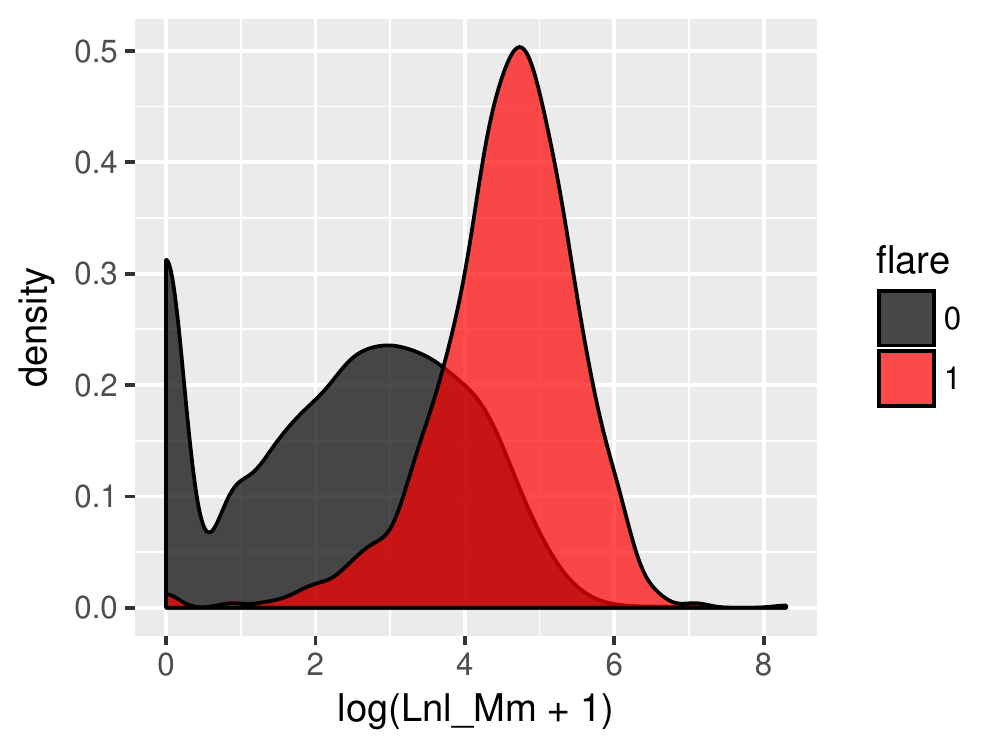}
}
\centerline{
\hspace*{0.015\textwidth}
\includegraphics[width=7cm,height=5cm]{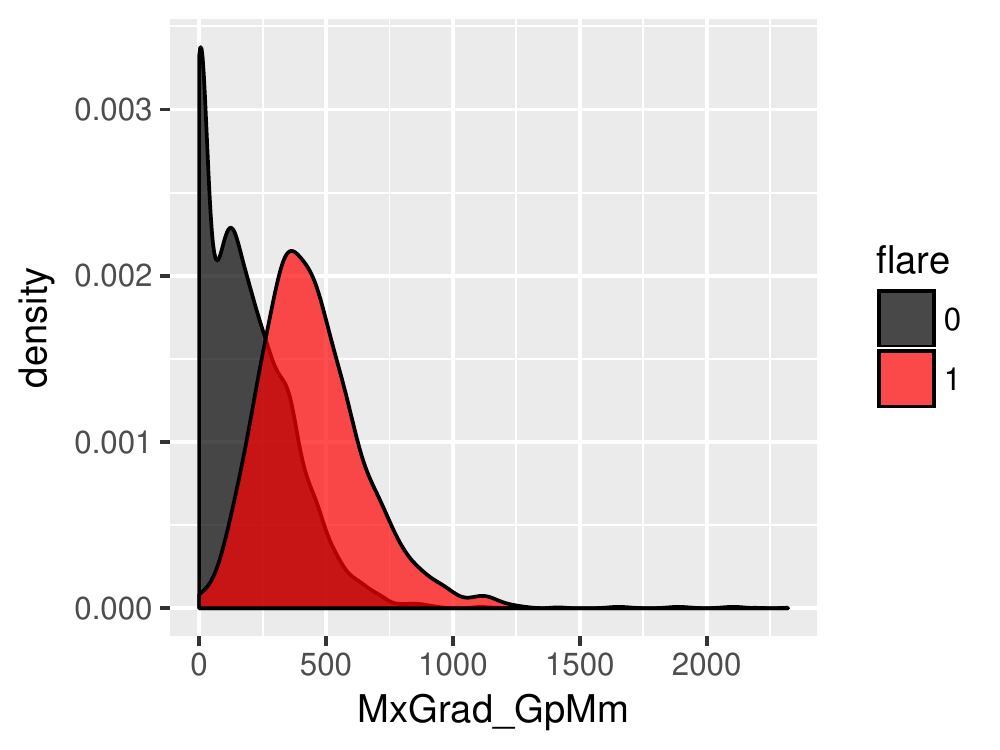}
\hspace*{-0.03\textwidth}
\includegraphics[width=7cm,height=5cm]{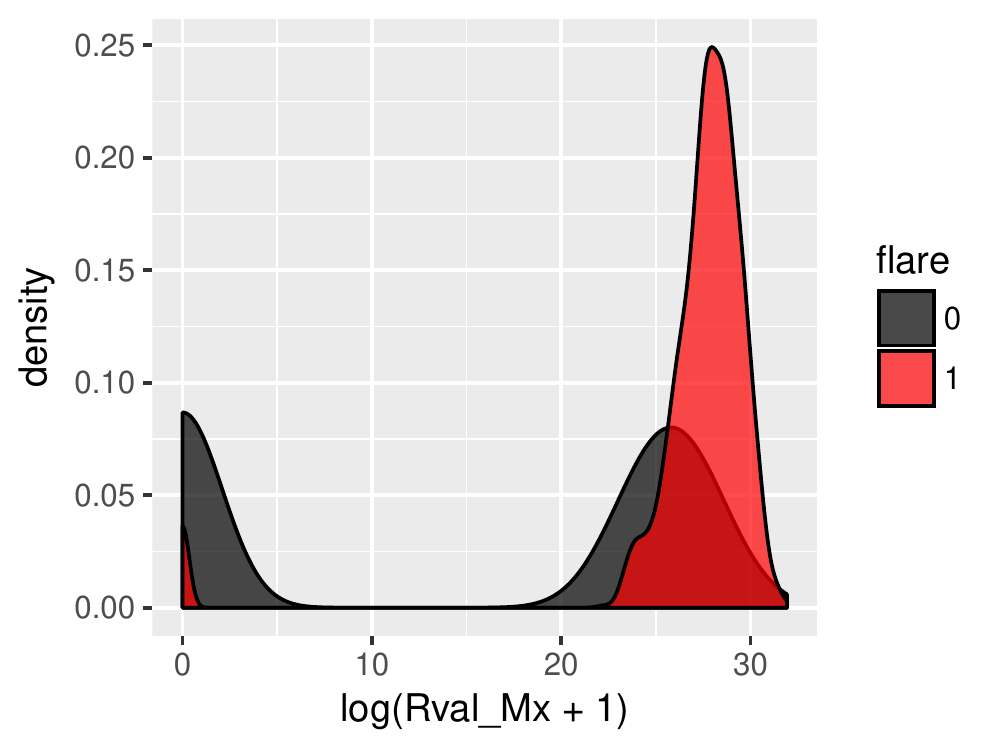}
}

\centerline{
\hspace*{0.015\textwidth}
\includegraphics[width=7cm,height=5cm]{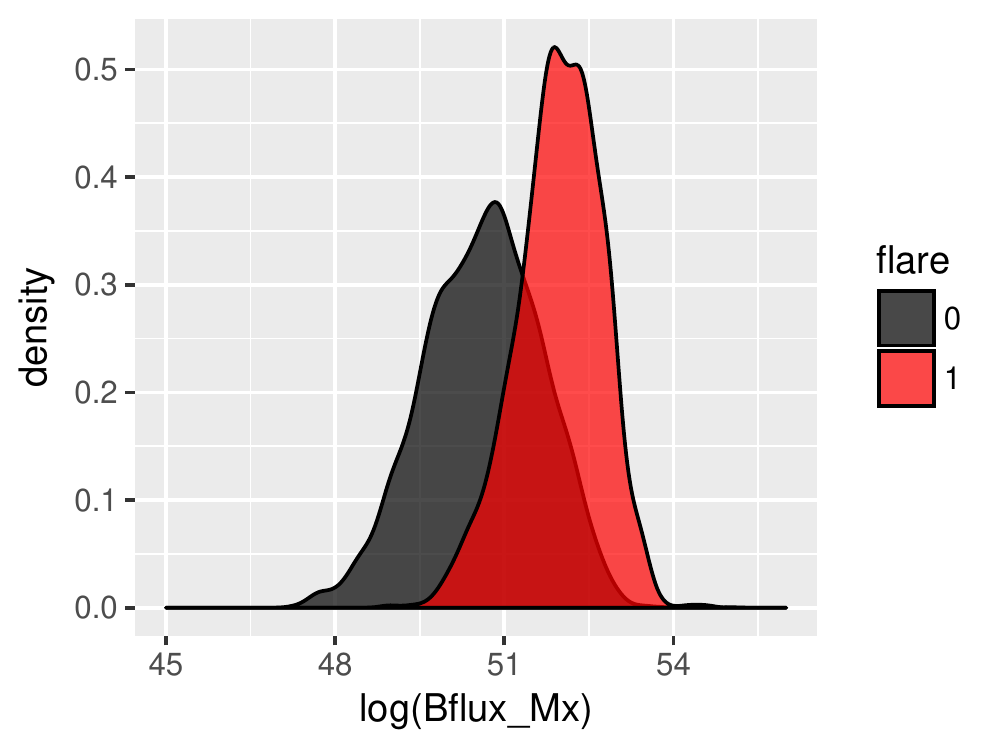}
\hspace*{-0.03\textwidth}
\includegraphics[width=7cm,height=5cm]{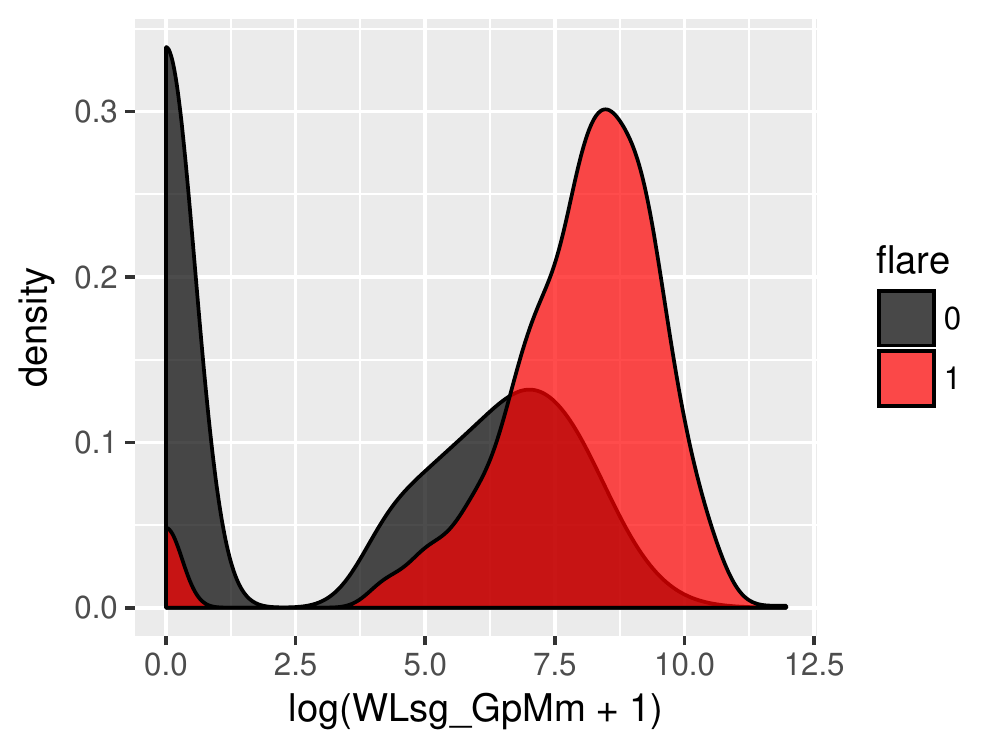}
}
  \caption{The marginal densities of the top features in the NOAA AR dataset selected by MR, Lasso and RF algorithms.}\label{fig:topfeats2}
\end{figure}

\begin{table}[t]
\caption{Results from the full SMART dataset for logistic regression 3 features (LR): median values for TSS, TPR, TNR, ACC and HSS for the testing set. 2.5th and 97.5th percentiles are given in brackets. The classification rules were obtained from 50 randomly sampled training subsets of 400 instances each. }\label{table:test}
\begin{tabular}{l|cc|ccc|cc}
  \hline
p & TSS & & TPR & TNR & ACC & HSS & \\
  \hline
0.01 & 0.8 & (0.79, 0.82) & 0.98 & 0.82 & 0.83 & 0.34 & (0.31, 0.37) \\
  0.02 & 0.82 & (0.81, 0.83) & 0.97 & 0.85 & 0.86 & 0.38 & (0.35, 0.42) \\
  0.03 & 0.83 & (0.82, 0.84) & 0.97 & 0.87 & 0.87 & 0.41 & (0.38, 0.45) \\
  0.04 & 0.84 & (0.83, 0.84) & 0.96 & 0.88 & 0.88 & 0.43 & (0.4, 0.47) \\
  0.05 & 0.84 & (0.83, 0.84) & 0.95 & 0.89 & 0.89 & 0.46 & (0.41, 0.49) \\
  0.06 & 0.84 & (0.83, 0.84) & 0.94 & 0.9 & 0.9 & 0.47 & (0.43, 0.51) \\
  0.07 & 0.84 & (0.83, 0.84) & 0.93 & 0.91 & 0.91 & 0.49 & (0.45, 0.52) \\
  0.08 & 0.84 & (0.83, 0.84) & 0.92 & 0.91 & 0.91 & 0.5 & (0.46, 0.54) \\
  0.09 & 0.83 & (0.83, 0.84) & 0.92 & 0.92 & 0.92 & 0.52 & (0.48, 0.55) \\
  0.1 & 0.83 & (0.82, 0.84) & 0.91 & 0.92 & 0.92 & 0.53 & (0.49, 0.56) \\
  0.2 & 0.78 & (0.74, 0.82) & 0.83 & 0.95 & 0.95 & 0.61 & (0.57, 0.64) \\
  0.3 & 0.69 & (0.61, 0.77) & 0.72 & 0.97 & 0.96 & 0.63 & (0.6, 0.64) \\
  0.4 & 0.58 & (0.5, 0.71) & 0.6 & 0.98 & 0.96 & 0.61 & (0.58, 0.64) \\
  0.5 & 0.49 & (0.38, 0.65) & 0.5 & 0.99 & 0.96 & 0.57 & (0.5, 0.63) \\
  0.6 & 0.37 & (0.26, 0.57) & 0.38 & 0.99 & 0.96 & 0.49 & (0.38, 0.61) \\
  0.7 & 0.26 & (0.15, 0.48) & 0.26 & 1 & 0.96 & 0.38 & (0.24, 0.57) \\
  0.8 & 0.14 & (0.06, 0.36) & 0.14 & 1 & 0.95 & 0.24 & (0.11, 0.48) \\
  0.9 & 0.04 & (0, 0.19) & 0.04 & 1 & 0.95 & 0.07 & (0.01, 0.31) \\
   \hline
\end{tabular}

\end{table}

\begin{table}[t]
\caption{Results from the full SMART dataset for logistic regression with 3 features trained on the full training dataset:  TSS, TPR, TNR, ACC and HSS for the testing set.}\label{table:testLRfull}
\begin{tabular}{rrrrrr}
  \hline
p & TSS & TPR & TNR & ACC & HSS \\
  \hline
0.01 & 0.80 & 0.98 & 0.82 & 0.83 & 0.33 \\
0.03 & 0.83 & 0.97 & 0.87 & 0.87 & 0.41 \\
0.05 & 0.84 & 0.95 & 0.89 & 0.89 & 0.46 \\
0.07 & 0.84 & 0.93 & 0.91 & 0.91 & 0.49 \\
0.09 & 0.84 & 0.92 & 0.92 & 0.92 & 0.52 \\
0.10 & 0.83 & 0.91 & 0.92 & 0.92 & 0.53 \\
0.20 & 0.79 & 0.83 & 0.96 & 0.95 & 0.62 \\
0.30 & 0.70 & 0.72 & 0.97 & 0.96 & 0.64 \\
0.40 & 0.60 & 0.62 & 0.98 & 0.96 & 0.62 \\
0.50 & 0.50 & 0.51 & 0.99 & 0.96 & 0.59 \\
0.60 & 0.39 & 0.40 & 0.99 & 0.96 & 0.52 \\
0.70 & 0.27 & 0.27 & 1.00 & 0.96 & 0.39 \\
0.80 & 0.14 & 0.14 & 1.00 & 0.95 & 0.24 \\
0.90 & 0.04 & 0.04 & 1.00 & 0.95 & 0.07 \\
   \hline
\end{tabular}

\end{table}

\begin{table}[t]
\caption{Results from the NOAA AR dataset for logistic regression 3 features (LR): median values for TSS, TPR, TNR, ACC and HSS for the testing set. 2.5th and 97.5th percentiles are given in brackets. The classification rules were obtained from 50 randomly sampled training subsets of 400 instances each. }\label{table:testNOAA}
\begin{tabular}{l|cc|ccc|cc}
  \hline
p & TSS & & TPR & TNR & ACC & HSS & \\
  \hline
  0.1 & 0.59 & (0.57, 0.63) & 0.93 & 0.65 & 0.71 & 0.4 & (0.37, 0.45) \\
  0.2 & 0.66 & (0.65, 0.66) & 0.86 & 0.8 & 0.81 & 0.53 & (0.5, 0.55) \\
  0.3 & 0.63 & (0.6, 0.65) & 0.75 & 0.88 & 0.85 & 0.58 & (0.56, 0.59) \\
  0.4 & 0.57 & (0.5, 0.61) & 0.64 & 0.93 & 0.87 & 0.58 & (0.55, 0.59) \\
  0.5 & 0.47 & (0.37, 0.55) & 0.51 & 0.96 & 0.87 & 0.53 & (0.45, 0.58) \\
  0.6 & 0.34 & (0.24, 0.47) & 0.37 & 0.98 & 0.85 & 0.43 & (0.32, 0.53) \\
  0.7 & 0.22 & (0.12, 0.35) & 0.23 & 0.99 & 0.84 & 0.31 & (0.18, 0.44) \\
  0.8 & 0.11 & (0.04, 0.24) & 0.11 & 1 & 0.82 & 0.16 & (0.07, 0.33) \\
  0.9 & 0.02 & (0, 0.11) & 0.02 & 1 & 0.8 & 0.04 & (0, 0.16) \\
   \hline
\end{tabular}
\end{table}

\begin{table}[ht]
\caption{ TSS, TPR, TNR, ACC and HSS for LR13 for the NOAA AR  dataset.}\label{table:testDNNfull}
\begin{tabular}{rrrrrr}
\hline
p & TSS & TPR & TNR & ACC & HSS \\
\hline
  0.10 & 0.60 & 0.93 & 0.67 & 0.72 & 0.41 \\
  0.20 & 0.67 & 0.87 & 0.80 & 0.81 & 0.54 \\
  0.30 & 0.63 & 0.76 & 0.87 & 0.85 & 0.57 \\
  0.40 & 0.57 & 0.66 & 0.92 & 0.86 & 0.58 \\
  0.50 & 0.52 & 0.57 & 0.95 & 0.87 & 0.56 \\
  0.60 & 0.41 & 0.44 & 0.97 & 0.86 & 0.49 \\
  0.70 & 0.30 & 0.31 & 0.98 & 0.85 & 0.39 \\
  0.80 & 0.17 & 0.17 & 1.00 & 0.83 & 0.24 \\
  0.90 & 0.04 & 0.04 & 1.00 & 0.81 & 0.07 \\
   \hline
\end{tabular}
\end{table}

\bibliographystyle{spr-mp-sola}


\end{article} 

\end{document}